\documentclass[12pt,preprint]{aastex}

\def\rhoc{$\rho_c$}
\def\deltav{$\Delta_v$}
\def\fgas{$f_{gas}$}
\def\mgas{$M_{gas}$}
\def\mtot{$M_{tot}$}
\def\nclu{38}

\newcommand{\Ho}{\mbox{$H_0$}}
\newcommand{\no}{\mbox{$n_{e0}$}}

\newcommand{\dTo}{\mbox{$\Delta T_0$}}
\newcommand{\Xo}{\mbox{$S_{x  0}$}}

\newcommand{\Msun}{\mbox{M$_\odot$}}

\newcommand{\Da}{\mbox{$D_{\!\mbox{\tiny A}}$}}

\newcommand{\lsim}{\lesssim}

\newcommand{\sze}{SZE}
\newcommand{\chandra}{{\it Chandra}}

\newcommand{\asca}{{\it ASCA}}

\newcommand{\bima}{{\it BIMA}}
\newcommand{\ovro}{{\it OVRO}}

\newcommand{\Mgas}{\mbox{$M_{\mbox{\scriptsize gas}}$}}
\newcommand{\Mtot}{\mbox{$M_{\mbox{\scriptsize tot}}$}}

\newcommand{\Ysz}{\mbox{$Y$}}
\newcommand{\rtfh}{\mbox{$r_{\mbox{\scriptsize 2500}}$}}

\def\sza{\it Sunyaev-Zeldovich Array\rm}
\begin{document}

\title{Scaling Relations
from Sunyaev-Zel'dovich Effect and Chandra X-ray measurements
of high-redshift galaxy clusters}

\author{
Massimiliano Bonamente\altaffilmark{1,2},
Marshall Joy\altaffilmark{2},
Samuel J. LaRoque\altaffilmark{3},
John E. Carlstrom\altaffilmark{3,4},
Daisuke Nagai\altaffilmark{5} and
Daniel P. Marrone\altaffilmark{6,3}
}

\altaffiltext{1}{Department of Physics, University of Alabama,
  Huntsville, AL 35812}

\altaffiltext{2}{NASA/Marshall Space Flight Center, Huntsville, AL 35812}

\altaffiltext{3}{Kavli Institute for Cosmological Physics, Department
  of Astronomy and Astrophysics, University of Chicago, Chicago, IL 60637}

\altaffiltext{4}{Department of Physics, Enrico Fermi Institute, University of Chicago,
  Chicago, IL 60637}

\altaffiltext{5}{Theoretical Astrophysics, California Institute of Technology, Mail Code 130-33, Pasadena, CA 91125}

\altaffiltext{6}{Jansky Fellow, National Radio Astronomy Observatory}

\begin{abstract}

We present Sunyaev-Zel'dovich Effect (SZE) scaling relations for
\nclu\ massive galaxy clusters at redshifts $0.14 \leq z \leq 0.89$,
observed with both the \chandra\ X-ray Observatory and the
centimeter-wave SZE imaging system at the BIMA and OVRO
interferometric arrays.  An isothermal $\beta$-model with central
100~kpc excluded from the X-ray data is used to model the intracluster
medium and to measure global cluster properties. For each cluster, we
measure the X-ray spectroscopic temperature, SZE gas mass, total mass
and integrated Compton-$y$ parameters within \rtfh.  Our measurements
are in agreement with the expectations based on a simple self-similar
model of cluster formation and evolution. We compare the cluster
properties derived from our SZE observations with and without
\chandra\ spatial and spectral information and find them to be in good
agreement. We compare our results with cosmological numerical
simulations, and find that simulations that include radiative cooling,
star formation and feedback match well both the slope and
normalization of our SZE scaling relations.

\end{abstract}

\section{Introduction}
\label{sec:intro}

The Sunyaev-Zel'dovich Effect (SZE) is a unique and powerful
observational tool for cosmology (for review see \citealt{carlstrom2002}). It
is a small distortion in the cosmic microwave background (CMB)
spectrum caused by scattering of CMB photons off a distribution of
high energy electrons in dense structures such as clusters of galaxies
\citep{sunyaev1970,sunyaev1972}.  This effect has a unique property
that the signal is independent of redshift, making it particularly
well suited for deep cluster surveys
\citep[e.g.,][]{holder2000,weller2002}.  Several \sze\ survey
experiments are currently in progress
\citep{Ruhl2004,Fowler2004,Kaneko2006}, and are expected to generate a
large sample of SZ-selected clusters with masses greater than $\sim
2\times10^{14}\,\Msun$.  The resulting  large samples of galaxy clusters will
enable direct measurements of the evolution of the number density of
galaxy clusters as a function of redshift and in principle can provide a 
powerful constraint on the nature of dark energy
\citep{wang1998,viana1999,mohr2000b,haiman2001}.

To utilize the upcoming SZE cluster surveys for cosmological studies,
it is important to understand the relation between the SZE observables
and the mass of a cluster. If the evolution of clusters is dominated
by gravitational processes, a simple model of cluster formation and
evolution based on the virial theorem \citep{kaiser1986} predicts
simple power-law relations between cluster masses and certain
integrated cluster properties, including the integrated SZE flux
(which is proportional to \Ysz, the integral of the Compton-$y$
parameter over the solid angle of the cluster). Numerical simulations
further suggest that \Ysz\ should be an excellent proxy of cluster
mass when measured on sufficiently large scales
\citep[e.g.,][]{dasilva2004,motl2005,nagai2006}. These simulations
also predict that the slope and redshift evolution of the SZE scaling
relations are relatively insensitive to the details of cluster
physics, although numerical simulations show that the input cluster
physics affects the normalization of the SZE scaling relations
\citep{nagai2006}. It is therefore important to investigate the
properties of the SZE scaling relations observationally.

Previous studies have addressed the correlation between the SZE signal
and X-ray properties.  For instance, \citet{cooray1999} found a
positive correlation between the central SZE decrement and the X-ray
luminosity in a sample of 14 clusters.  Similarly,
\citet{mccarthy2003} detected correlations between the central SZE
decrement and X-ray determined mass, temperature, and luminosity for a
22 cluster sample, and \citet{morandi2007}
for a sample of 24 clusters.  These studies use data from multiple SZE and X-ray
experiments, making systematics more difficult to control, and focus
on the relationship between the central values of the SZE signal with
the X-ray properties.  Recently, \citet{benson2004} showed that
the integrated SZE flux is a more robust observable than the central
values of the SZE signal, and found a strong correlation with X-ray
temperatures using a sample of 15 clusters obtained by the
Sunyaev-Zeldovich Imaging Experiment (SuZIE,
\citealt{holzapfel1997a,benson2003}) and X-ray temperatures from the
\asca\ experiment.

This paper is the third in a series of papers combining SZE and
\chandra\ X-ray measurements of galaxy clusters to study cosmological
properties, following \citet{bonamente2006} (B2006 hereafter) and
\citet{laroque2006} (L2006 hereafter).  Here we present observational
studies of SZE scaling relations for clusters of galaxies.  This paper
advances the results of previous cluster scaling relation works in
several ways.  First, we use the largest observational sample yet
constructed (\nclu\ clusters at redshift $z$=0.14---0.89).  Second,
our analysis is based on SZE and X-ray observations obtained using the
same instruments: all SZE fluxes are determined using centimeter-wave
interferometric data from the BIMA/OVRO SZE imaging experiments
\citep[e.g.][]{laroque2003}, and all cluster X-ray properties are
derived using data from the \chandra\ {\it X-ray Observatory}.
Finally, our \chandra\ observations have an order of magnitude better
spatial resolution than the X-ray data used in previous studies, which
greatly improves our ability to identify and exclude compact
foreground sources which are superimposed on the cluster X-ray
emission.

Throughout the paper, we assume a $\Lambda$CDM cosmology with 
$\Omega_{M}$=0.3, $\Omega_{\Lambda}$=0.7 and $h=0.7$, where $h$ is defined such that
$\Ho=100\,h$~km~s$^{-1}$~Mpc$^{-1}$.  All uncertainties are at the
68.3\% confidence level.

\section{Theory of cluster scaling relations}
\label{theory}
\subsection{The virial radius and \rtfh}
In order to establish relationships between mass, SZE flux and other
cluster properties, one needs to define a radius out to which all
quantities will be calculated.  This radius should be physically
motivated, reachable with the current X-ray and SZE observations, and
equivalent for clusters of different redshift.  One candidate is the
virial radius.  In a Friedman-Robertson-Walker universe, an
unperturbed spherical region expands indefinitely, while a perturbed
overdense region (the seed of a future cluster) eventually
recollapses.  When the overdense region collapses under the effect of
its own gravity, it is assumed to reach virial equilibrium when the
radius is half of that at maximum expansion
\citep{peebles1980,lacey1993}.  The ratio of the mean cluster density
to the background density at the time of virialization is $\Delta_v=18
\pi^2$ for a universe with critical matter density ($\Omega_{M}=1$).
For a different cosmology with $\Omega_k=0$, \citet{bryan1998} showed
that $\Delta_v \simeq 18 \pi^2+82 x -39 x^2$, where $x=\Omega_{M0}
(1+z)^3/E^2(z)$ and $E^2(z)=\Omega_{M0}
(1+z)^3+\Omega_{\Lambda}+\Omega_{k0} (1+z)^2$, as found from a fit to
numerical simulations \citep{lacey1993}.

With this characterization of the mean cluster density at time of
virialization, the virial radius can be determined as the radius
within which the average density of the cluster is \deltav\ times
the critical density, via

\begin{equation}
\frac{4}{3} \pi   \rho_c(z)  \Delta_v(z)   r_{vir}^3 = M_{tot}(r_{\Delta_v(z)})
\label{rvir}
\end{equation}
in which both \rhoc($z$) and \deltav($z$) are  cosmology
dependent, and the critical density $\rho_c(z)$ is defined as:
\begin{equation}
\rho_c(z)=\frac{3   H_0^2 E(z)^2}{8 \pi G} .
\label{rhocrit}
\end{equation}

Unfortunately, the virial radius is usually unreachable with current
X-ray and SZE measurements, and one is forced to perform measurements
out to a smaller radius.  Such a radius ($r_{\Delta}$) is
characterized by the density contrast parameter $\Delta$ in place of
$\Delta_v(z)$ in Equation \ref{rvir}, and corresponds to a higher
average density, ${4}/{3} \cdot \pi \rho_c(z) \Delta \cdot
r_{\Delta}^3 = M_{tot}(r_{\Delta})$.  We choose a contrast parameter
$\Delta=2500$, corresponding to an average density of 2500 times the
critical density at the cluster's redshift. This choice is motivated
by the fact that this is the radius typically reachable with our \sze\
and X-ray data without any extrapolation of the models (B2006,
L2006).~\footnote{The use of a constant overdensity factor $\Delta$
was shown by \citet{maughan2006} to give results similar to the case
of a variable overdensity factor
$\Delta(z)=\Delta(0)[\Delta_v(z)/\Delta_v(0)]$, in which the variable
overdensity scales with redshift in order to keep the ratio of two
comoving densities constant.}

\subsection{Scaling relations \label{scaling:theory}}
The hierarchical structure formation theory developed by
\citet{kaiser1986} predicts simple relationships between physical
parameters of collapsed structures, known as scaling relations.  With
the assumptions of hydrostatic equilibrium and of an isothermal
distribution for both the dark matter and the cluster gas
\citep[e.g.,][]{bryan1998}, it can be shown that there is a simple
relationship between a cluster's total mass and its gas temperature
$T_e$:

\begin{equation}
T_e \propto M_{tot}^{2/3} E(z)^{2/3}
\label{m-t}
\end{equation}
where the mass is calculated out to a radius of mean overdensity
$\Delta$, $M_{tot}=M_{tot}(r_{\Delta})$.  For $f_{gas}~\equiv~M_{gas}/
M_{tot}$, ($M_{gas}=M_{gas}(r_{\Delta}$)), the expected relationship between the gas mass within
$r_\Delta$ and the gas temperature is

\begin{equation}
T_e f^{2/3}_{gas} \propto  M_{gas}^{2/3}  E(z)^{2/3} .
\label{mgas-t}
\end{equation}

The Compton-$y$ parameter is a measure of the pressure integrated
along the line of sight:
\begin{equation}
y=\int_0^{\infty} \sigma_T n_e \frac{k_B T_e}{m_e c^2} dl
\end{equation}
One can further integrate the $y$ parameter over the solid angle
$\Omega$ subtended by the cluster, to obtain the integrated
Compton-$y$ parameter:
\begin{equation}
Y\equiv\int_{\Omega} y d\Omega = 
\frac{1}{D_A^2} \left(\frac{k_B \sigma_T}{m_e c^2}\right) 
\int_0^{\infty} dl \int_A n_e T_e dA \\
\label{y-def}
\end{equation}
where $A$ is the area of the cluster in the plane of the sky.
In the context of an isothermal model, \Ysz\ is proportional to the
integral of the electron density $n_e$
over a cylindrical volume, thus
\begin{equation}
Y D_A^2 \propto T_e \int  n_e dV  = M_{gas} T_e = 
f_{gas} M_{tot} T_e .
\label{y-da}
\end{equation}
In section \ref{sz-x} we consider the effect of integrating gas mass
within a spherical volume while determining $Y$ in a cylinder.
Using equation \ref{m-t} we can rewrite equation \ref{y-da} in terms
of either $M_{tot}$ or $T_e$, or substitute $M_{gas}/f_{gas}$ for $M_{tot}$, to obtain:
\begin{eqnarray}
Y  D_A^2 \propto
f_{gas}  T_e^{5/2} E(z)^{-1} \label{Y-T} \nonumber\\
Y  D_A^2 \propto f_{gas}  M_{tot}^{5/3} E(z)^{2/3} \label{Y-Mtot}\\
Y  D_A^2 \propto f_{gas}^{-2/3}  M_{gas}^{5/3} E(z)^{2/3} \label{Y-Mgas} \nonumber
\end{eqnarray}
Equations \ref{Y-T} are the scaling 
relations that we investigate observationally in this paper.

\section{\sze\ and \chandra\ X-ray observations of galaxy clusters}
\label{data}

\subsection{Data}
We analyze the \sze\ and X-ray data observations of \nclu\ clusters in
the redshift range $z$=0.14---0.89, observed with the \bima\ and
\ovro\ interferometric arrays and with the \chandra\ X-ray imaging
spectrometers.  Both data modeling with the isothermal $\beta$ model
and the data themselves are presented in B2006 and L2006, the previous
two papers in this series.  We refer to L2006 for details on the
observations and data modeling, and to \citet{reese2002} for a
detailed illustration of the modeling of the OVRO/BIMA SZE data in the
Fourier plane.  In the following, we review those aspects of the data
modeling and analysis that are relevant to the investigation of the
scaling relations.

\subsection{Data modeling}
\label{model}
The gas density model is based on the spherical $\beta$-model
\citep{cavaliere1976, cavaliere1978}, which has the form
\begin{eqnarray}
n_e(r) = \no \left ( 1 + \frac{r^2}{r_c^2} \right )^{-3\beta/2} \nonumber ,
\end{eqnarray}
where $n_{e0}$ is the central electron number density, $r$ is the
radius from the center of the cluster, $r_c$ is a core radius, and
$\beta$ is a power-law index.  When integrated along the line of sight
to determine the projected
SZE decrement distribution ($\propto n_e$) and X-ray surface
brightness ($\propto n_e^2$), this model has the simple analytic forms
\begin{eqnarray}
\Delta T & = & \dTo \left ( 1 + \frac{\theta^2}{\theta_c^2} \right)
^{(1-3\beta)/2} \label{eq:easy_sz_signal}\\
S_x & = & \Xo \left ( 1 + \frac{\theta^2}{\theta_c^2} \right
        )^{(1-6\beta)/2}, \label{eq:easy_x_signal}
\end{eqnarray}
where \dTo\ is the central thermodynamic SZE temperature
decrement/increment and $\theta_c$ is the angular core radius of the
cluster \citep[e.g.,][]{birkinshaw1991,reese2002}.  This model
typically provides a good description of the X-ray surface brightness
and SZE decrement profiles out to $\sim$\rtfh\
\citep[e.g.,][]{jones1984,elbaz1995,grego2001,reese2002,ettori2004}.
This simple model, however, does not provide a good description of the
peaked X-ray surface brightness observed in the center of some
clusters.  To minimize the systematic bias associated with modeling of
cluster cores, we therefore exclude the central 100~kpc from both the
spatial and spectral X-ray data, as was done in the previous two
papers in this series.  The emission-weighted X-ray spectroscopic
temperature is also determined by a single-temperature fit to the
X-ray spectrum of photons extracted from an annulus between 100~kpc
and $r_{2500}$ (L2006).  This 100~kpc-cut model was shown to recover
the gas masses of simulated clusters with a range of dynamical states
to better than 5\% accuracy at \rtfh\ (L2006) and, when applied to the
determination of the Hubble constant, yielded the same results as a
more complex non-isothermal model (B2006).  Uncertainties associated
with the isothermal assumption are included via an additional
systematic error described in section \ref{method}.

Following our earlier analysis methods,
we do not model the dark matter distribution, and calculate the
total mass directly from the equation of hydrostatic equilibrium (see Equation
\ref{eq:mtot_hse_iso} below). An upcoming paper (Mroczkowski et al. 2008, in preparation)
will extend the analysis to use
more sensitive data from the \sza\ \citep{muchovej2007} and more accurate modeling of the
cluster gas, based on the non-isothermal models of \citet{vikhlinin2006}
and \citet{nagai2007b}.

\subsection{Analysis Methods}
\label{subsec:analysis}

Best-fit model parameters and confidence intervals for all model
parameters are obtained using a Markov chain Monte Carlo (MCMC) method
described in detail by \citet{bonamente2004} and B2006.  L2006
explains the implementation of the likelihood calculation for the
100~kpc-cut model.  For each cluster, the Markov chain constrains the
parameters $S_{x0}$, $\beta$, $\theta_c$, $\Delta T_0$, $T_e$, and
abundance (see L2006 for best-fit values).  We use the
cosmological parameters $h=0.7$, $\Omega_M$=0.3 and $\Omega_{\Lambda}$=0.7 to calculate each
cluster's angular diameter distance $D_A$
\citep[e.g.,][]{carroll1992}.

From these model parameters we calculate \rtfh\ and $M_{tot}(r_{2500})$
through the hydrostatic equilibrium equation
\citep[e.g.,][]{grego2001},
\begin{equation}
M_{tot}(r)=\frac{3\beta kT_X}{G \mu m_p} \frac{r^3}{r_c^2 + r^2},
\label{eq:mtot_hse_iso}
\end{equation}
in which $\mu$ is the mean molecular weight calculated using the X-ray
metallicities and $r_c=\theta_c D_A$. One obtains \rtfh\ from the solution of the
following equation:
\begin{eqnarray}
\frac{3\beta kT_X}{G \mu m_p} \frac{r_{\Delta}^3}{r_c^2 +
r_{\Delta}^2}= \frac{4\pi}{3} r_{\Delta}^3 \Delta \rho_c(z) , \nonumber
\end{eqnarray}
in which $\rho_c(z)$ is given by Equation \ref{rhocrit},
$\Delta=2500$, and the right hand side is just $M_{tot}(r_{2500})$.
We then compute the global
cluster quantities needed for the analysis of SZE and X-ray scaling
relations. 
The gas mass is computed by integrating the gas density model,
\begin{equation}
\Mgas(r_{2500}) = 4\pi \mu_e \no m_p\, \Da^3 \int_{0}^{r_{2500}/\Da} \left
(1+\frac{\theta^2}{\theta_c^2} \right)^{-3\beta/2}\: \theta^2 d\theta,
\label{eq:mgas_single}
\end{equation}
where $\mu_e$ is the mean molecular weight of the electrons, and \no\
is the central electron density, obtained from the parameters of the
$\beta$ model (L2006, Equation 12).

The integrated $y$ parameter (\Ysz, equation \ref{y-def})
is calculated using the measured SZE decrement $\Delta T$, which is directly
proportional to the Compton-$y$ parameter
\begin{eqnarray}
\Delta T=T_{CMB} \cdot f(x) \cdot y \nonumber .
\end{eqnarray}
The factor $f(x)$ is the frequency dependence of the SZE:
\begin{eqnarray}
f(x) = \left(x \cdot coth \left(\frac{x}{2}\right) -4 \right) \cdot (1+\delta_{rel}) \nonumber ,
\end{eqnarray}
in which  $x=h \nu / k_B T_{CMB}$, and $\delta_{rel}$ is a small
relativistic correction factor. At our observing frequencies,
$f(x)\simeq -2$. Thus,
\begin{equation}
Y = \int_A \frac{\Delta T}{T_{CMB} f(x)} d\Omega=
\frac{\Delta T_o}{T_{CMB} f(x)} \int_0^{r_{2500}/D_A} 
\left( 1+\frac{\theta^2}{\theta_c^2}\right)^{(1-3\beta)/2} \theta d\theta .
\end{equation}

\section{Observational constraints on \sze\ scaling relations \label{scaling:obs}}

\subsection{Regression method \label{method}}
Our measurements of masses and integrated $y$ parameters, using the
method described in Section \ref{data}, are shown in Table
\ref{table:results}.  The errors in Table \ref{table:results}
represent the photon-counting statistical uncertainties of the X-ray
data and the statistical uncertainties of the SZE observations.
Additional sources of uncertainty in the measurement of cluster
parameters include cluster asphericity and projection effects,
small-scale clumping of the gas, the presence of point sources in the
field, CMB anisotropy, the assumption of isothermality, and
instrumental calibration, as discussed by \citet{reese2002}, L2006,
and B2006.  Therefore, in fitting \Ysz\ versus $T_X$, \mgas, and
\mtot\ (Equations \ref{Y-T}), we include an additional statistical
error (combined in quadrature) of $\pm 20$\% for the masses, and $\pm
10$\% for \Ysz\ and $T_X$.

We perform a linear least-squares regression in log space, 
$log(Y)=A+B \cdot log(X)$, following the
method of \citet{press1992} and \citet{benson2004}. This method accounts
for errors in both measured parameters for each scaling relation, and
it minimizes the $\chi^2$ statistic defined as
\begin{eqnarray}
\chi^2 = \sum \frac{(log(Y_i) -A -B log(X_i))^2}{\sigma^2_{log(Y_i)}+
(B \sigma_{log(X_i)})^2}  \nonumber
\end{eqnarray}
in which $\sigma_{log(Y_i)}=\sigma_{Y_i}/Y_i log(e)$, 
$\sigma_{log(X_i)}=\sigma_{X_i}/X_i log(e)$, and the linear errors
$\sigma_{Y_i}$ and $\sigma_{X_i}$ are obtained from the upper and lower 
uncertainties around the best-fit values as $\sigma=(\sigma^+ + \sigma^-)/2$.
                                                                                                            
\subsection{The \Ysz-\mgas, \Ysz-\mtot\ and \Ysz-$kT$ scaling relations \label{sec:scaling}}

\begin{figure}[!h]
\includegraphics[width=3.0in]{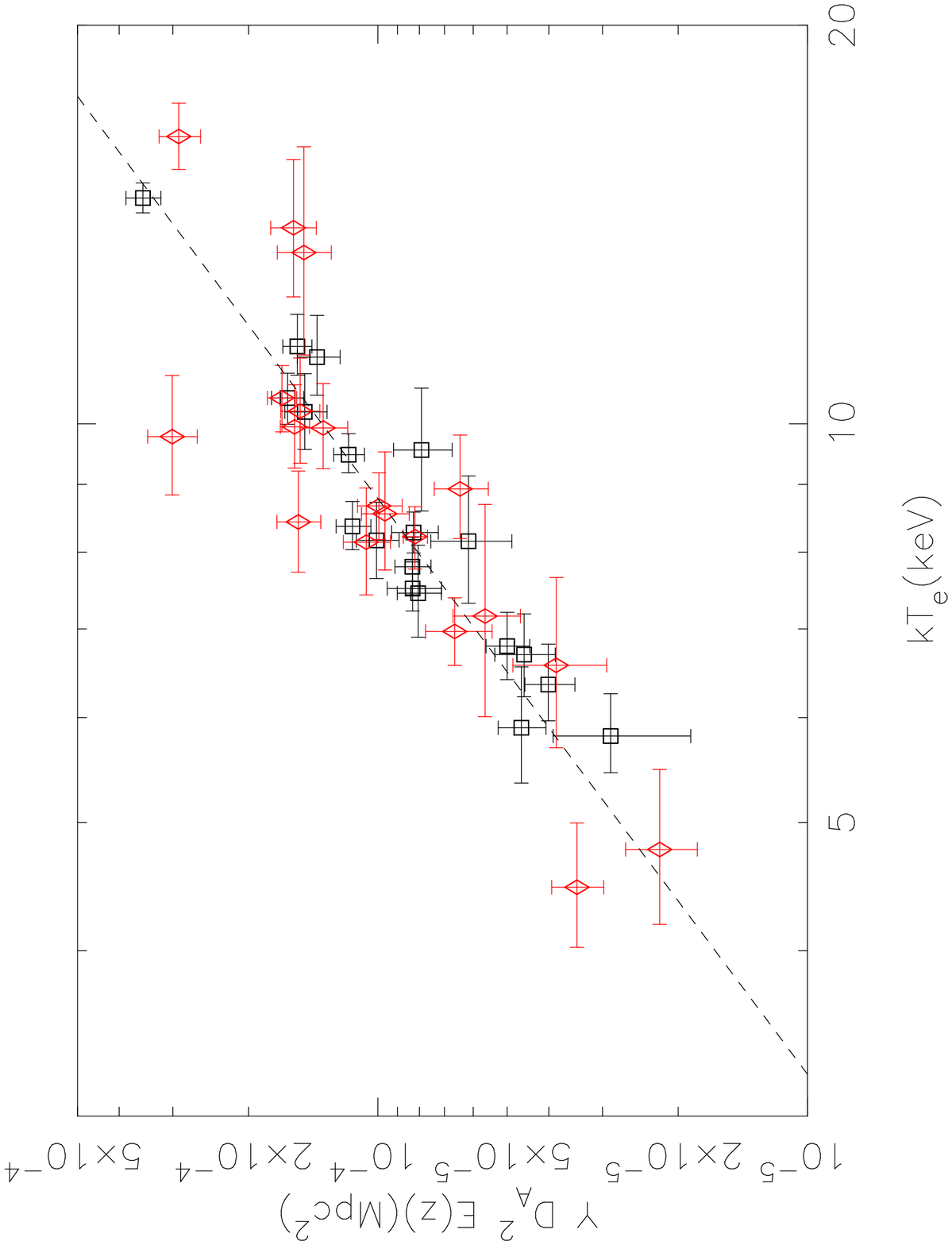}
\includegraphics[width=3.0in]{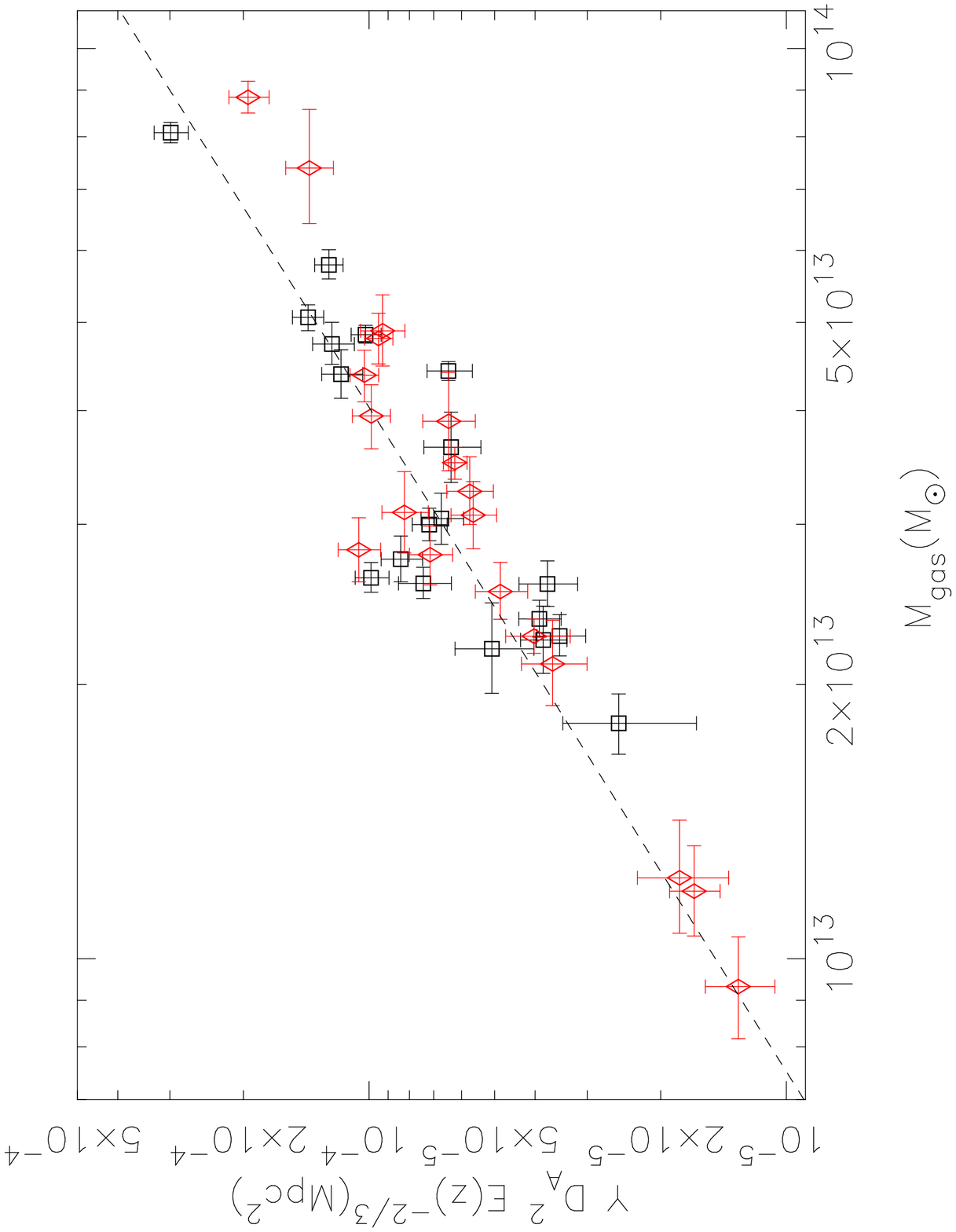}
\includegraphics[width=3.0in]{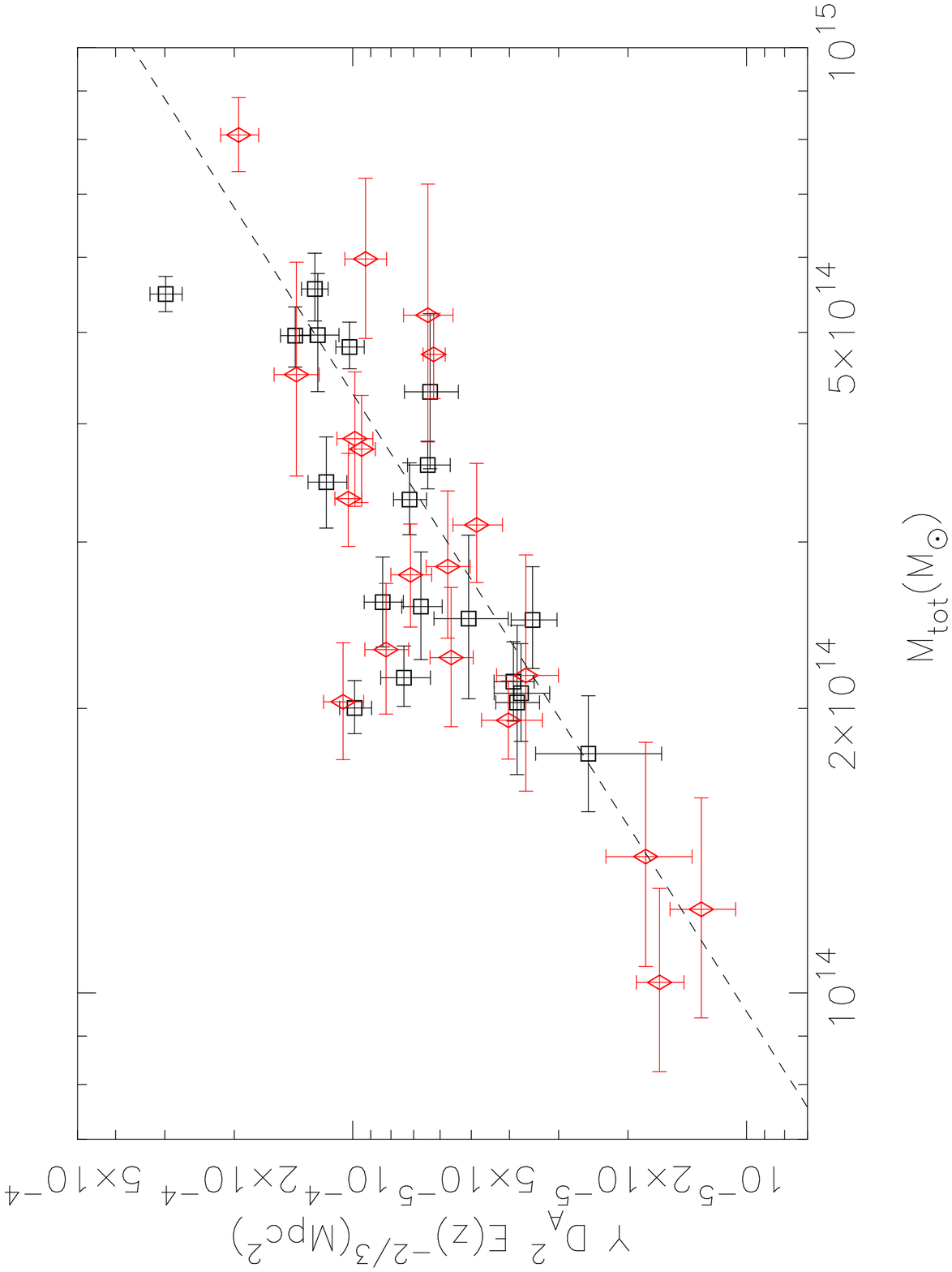}
\caption{Scaling relations between the \sze\ \Ysz\ 
and \mgas, \mtot\ and $kT_e$. Open squares (in black) are clusters at $0.14\leq z \leq 0.30$, 
open diamonds (in
red) are clusters at $0.30 < z \leq 0.89$. All measurements follow simple power-law models
with indices that are consistent with the values of the self-similar scaling theory
(Table \ref{table:results}).\label{fig:scaling}}
\end{figure}

The above derivation of the self-similar scaling relations does not
include any variation of the gas fraction with cluster mass. However,
there may be some evidence for such variation in both X-ray
observations (e.g., \citealt{vikhlinin2006}) and simulations
\citep{kravtsov2005}. We examine this in the present data
by performing a logarithmic fit
to the \fgas-\mgas\ data using a linear relationship ($log(Y)=A+B log(X)$).
We find no significant evidence for a variation of
\fgas\ with mass ($B=0.14\pm0.08$). In the following we therefore
assume that \fgas\ is a constant.
We then perform similar logarithmic fits to the \Ysz-\mgas, \Ysz-\mtot\ and \Ysz-$kT$ data.
The results
are shown in Figure \ref{fig:scaling} and in Table \ref{table:scaling}.
Under the assumption of a constant \fgas, all scaling relations are consistent
(within 2$\sigma$ statistical uncertainty)
with the simple self-similar model of cluster evolution.

\subsection{Redshift evolution of the \Ysz-\mgas, \Ysz-\mtot\ and \Ysz-$kT$  scaling relations}

The large number of clusters (38) and redshift coverage of our \chandra\ and \ovro-\bima\ 
data ($0.14\leq z \leq0.89$) enables
the investigation of a possible redshift evolution of the SZE scaling relations.
For this purpose, we divide our sample evenly into  low-redshift clusters ($z \leq 0.30$, 19 clusters)
and high-redshift clusters ($0.30< z \leq0.89$, 19 clusters), and repeat the 
logarithmic fits of Section \ref{sec:scaling}.

The results of Table \ref{table:scaling} indicate no evidence for
redshift evolution of the SZE scaling relations, as \fgas\ is
consistent with a constant for both low and high-redshift clusters.
Furthermore, the SZE scaling relations are consistent with the
self-similar slopes at or below the 2.5$\sigma$ level.

\section{Comparison of \sze\ and X-ray Measurements}
\label{sz-x}

In the previous section we have examined the \sze\ scaling relations
based on quantities derived jointly from SZE and X-ray
observations. Here we compare the cluster properties derived from our
SZE observations without using the X-ray data in the fits, in order to
determine whether the relations we observe depend strongly on the
X-ray information. 

We analyze the SZE data using the model of Section \ref{model}, using
additional assumptions to provide the constraints that would otherwise
be provided by the X-ray data. As a first assumption we fix
$\beta=0.7$ and fit for $\theta_c$ and $\Delta T_0$ in Equation
\ref{eq:easy_sz_signal}, following L2006.  The choice of fixing
$\beta=0.7$ is determined by the fact that this is the median value
for our sample; L2006 also show that using values of 0.6
and 0.8 results in changes to the parameters that are small relative
to the 68\% statistical uncertainties.  The data quality allows us to
perform this SZE-only analysis for 25 of the clusters in the full
sample, as shown in Table \ref{table:sz-results}. Knowledge of the gas
temperature is required in order to determine \rtfh, as can be seen
from the combination of Equations \ref{rvir} and
\ref{eq:mtot_hse_iso}:
\begin{eqnarray}
r_{2500}=\sqrt{\left( \frac{3 \beta k T}{G \mu m_p}\right) 
\frac{1}{\frac{4}{3} \pi \rho_c(z) \cdot 2500} -r_c^2}. \nonumber
\end{eqnarray}
In the absence of complementary X-ray spectroscopic data, we estimate
the gas temperature directly from the SZE data following the iterative
method described by \citet{joy2001}. We first choose an initial
estimate of the gas temperature, from which we obtain \rtfh\ and
$M_{tot}(r_{2500})$. We derive $M_{gas}(r_{2500})$ using the equations
described in Section \ref{subsec:analysis}, with the central gas
density $n_{e0}$ calculated from the parameters of the SZE decrement
model (L2006, Equation 13).  We provide a final constraint by assuming
that the gas mass fraction of each cluster is equal to the average
value for this sample, \fgas$=0.116$ (L2006), and iteratively solve
the equation $M_{gas}(r_{2500})=f_{gas} \cdot M_{tot}(r_{2500})$ in
order to find self-consistent estimates of $T$ and \rtfh.

\begin{figure}
\includegraphics[width=3.0in]{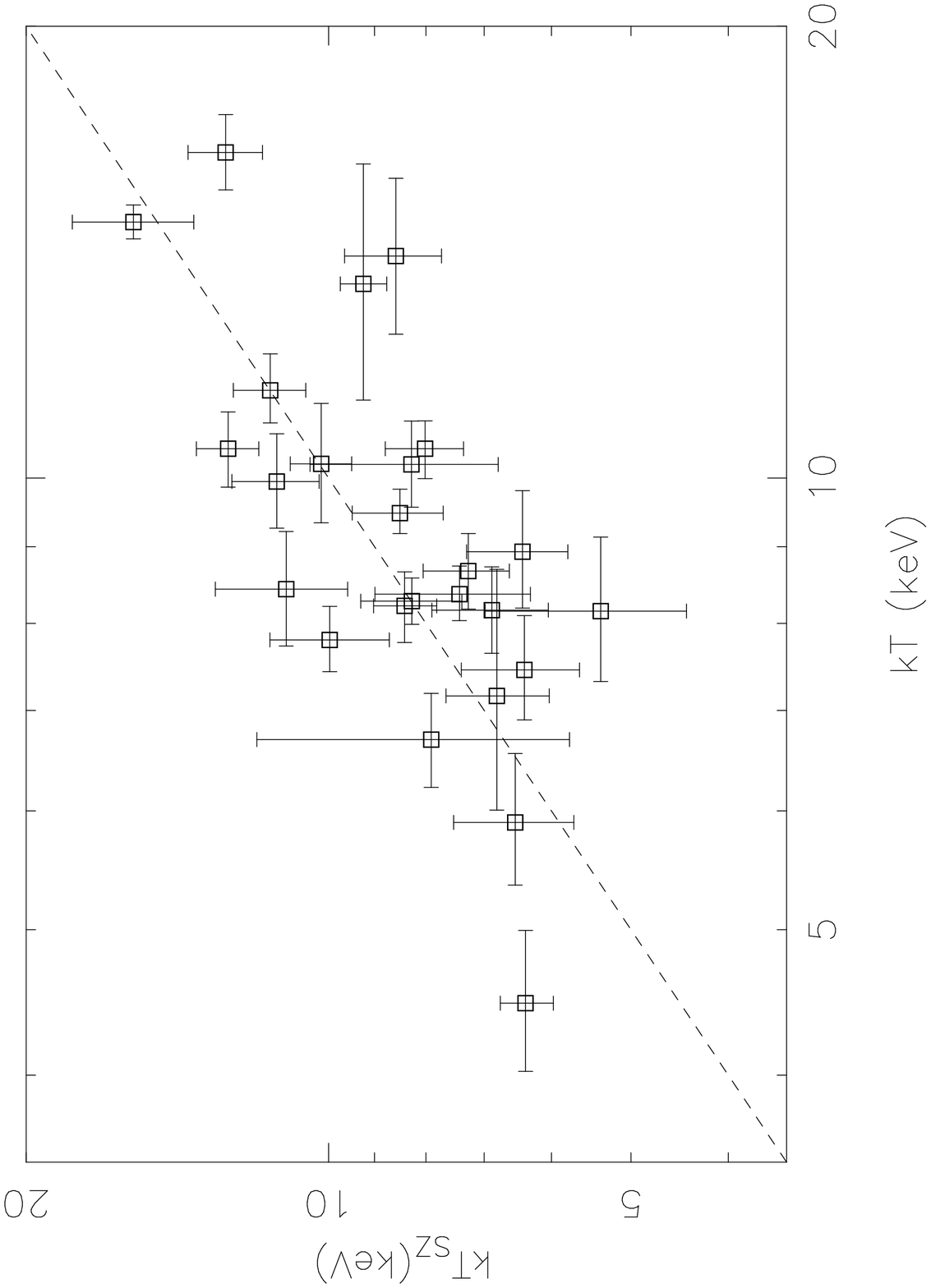}
\hspace{-0.0cm}
\includegraphics[width=3.0in]{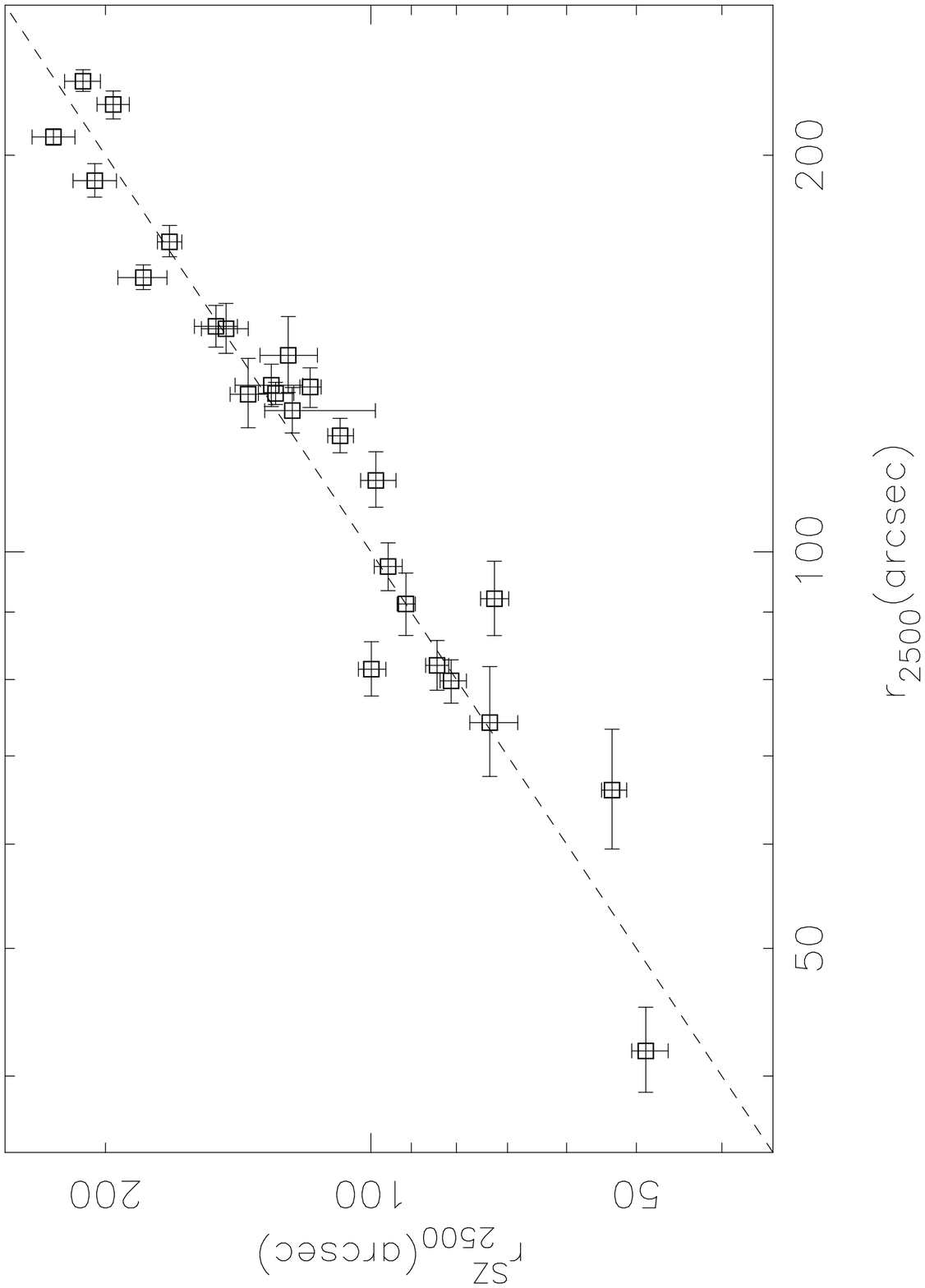}
\includegraphics[width=3.0in]{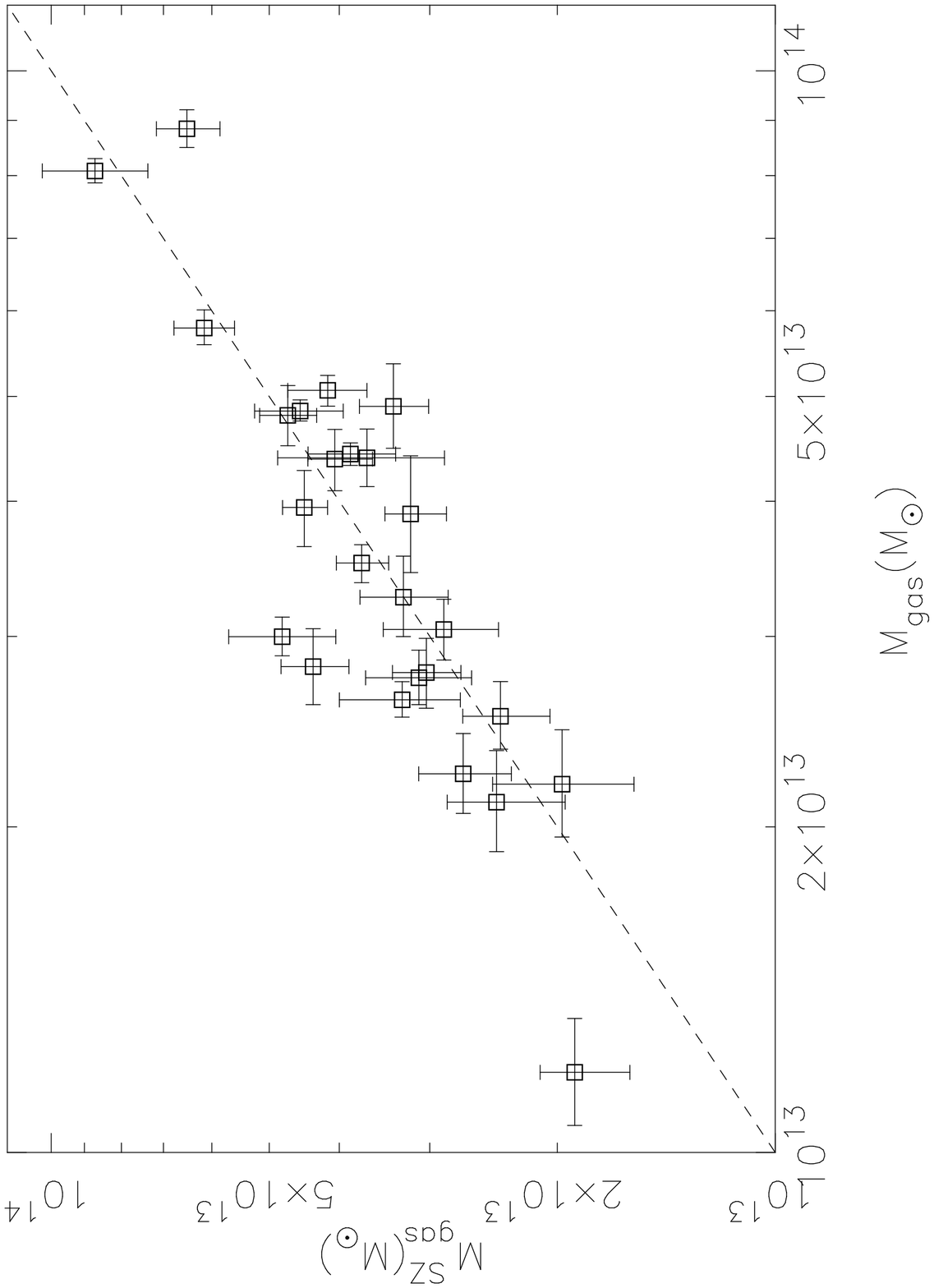}
\hspace{-0.0cm}
\includegraphics[width=3.0in]{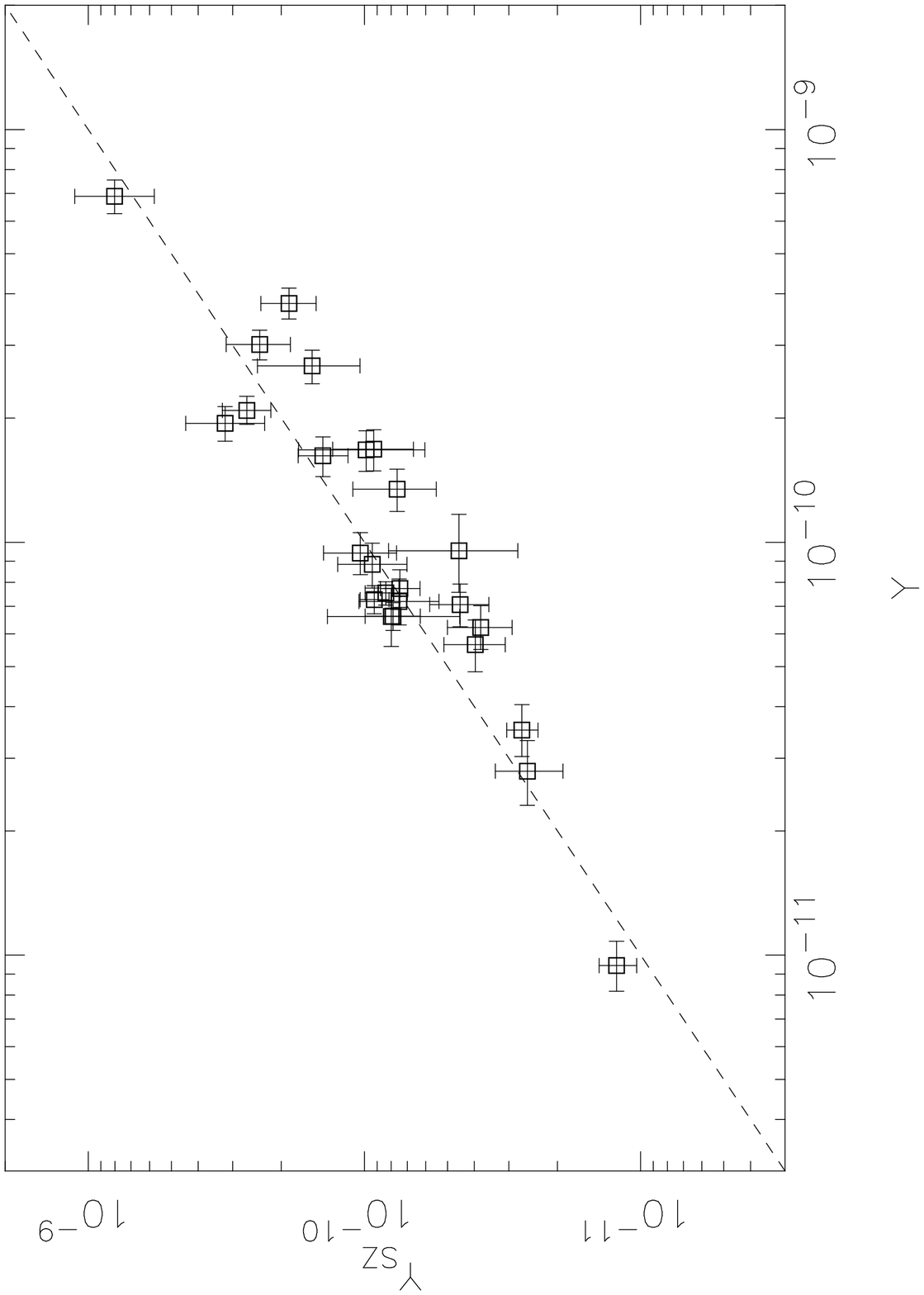}
\caption{Comparison between cluster parameters derived from joint analysis of
X-ray and SZE data ($x$ axis) and those derived from SZE data ($y$ axis)
following the procedure of Section~\ref{sz-x}. Dashed lines correspond to
$y=x$.\label{X-SZ-comparison}}
\end{figure}

Results of the SZE analysis are shown in Table
\ref{table:sz-results}. In Figure \ref{X-SZ-comparison} we compare the
SZE measurements of gas temperature, \rtfh, \Ysz, and gas mass with
the values from the joint analysis of Section \ref{data}.  In the
joint analysis, the X-ray data are solely responsible for the
measurement of the temperature, and drive the fit of the spatial
parameters and gas density.  They therefore drive the measurements of
$r_{2500}$ and $M_{gas}$ as well.  The quantities inferred from the
SZE data are in good agreement with those from the joint analysis,
indicating that we have not altered the scaling relations of Section
\ref{sec:scaling} by incorporating the X-ray data. Although these
results show that it is possible to estimate \Ysz\ and the cluster gas
mass from the SZE data alone, the joint analysis is preferred because
it provides the strongest constraints on cluster properties, requires
fewer assumptions about the cluster structure and composition, and can
be applied to the full 38-cluster sample. Nevertheless, the ability to
derive cluster properties from SZE observations
will be important for the many \sze\ cluster
surveys currently underway \citep{Ruhl2004,Fowler2004,Kaneko2006};
these surveys are expected to generate large samples of SZE-selected
clusters but will not generally have access to deep X-ray observations
for cluster characterization.  The gas temperature may also be
inferred from multi-frequency SZE observations
\citep{hansen2002}.


Recent work by \citet{kravtsov2006} indicate that the quantity $Y_X
\equiv M_{gas} \cdot k_B T_e$, the X-ray analogue of \Ysz, is a low-scatter proxy for the cluster total mass.  For the
isothermal $\beta$-model used in this paper, the gas mass can be
estimated from X-ray data by using Equation \ref{eq:mgas_single}, and
thus our data also provide a measurement of $Y_X$. We can
therefore compare $Y$ with $Y_X$, in order to establish
observationally whether the two quantities are indeed equivalent.  In the
case of the isothermal $\beta$-model, the integrated Compton-$y$ parameter is
an integral of the electron density over a cylinder $\mathcal{C}$ of
infinite length along the line of sight, and of area $A= \pi
r_{2500}^2$:
\begin{eqnarray}
Y D_A^2=\left(\frac{k_B \sigma_T T_e}{m_e c^2}\right)
\int_{\mathcal{C}} n_e(r) dV = \left(\frac{k_B \sigma_T T_e}{m_e c^2}\right) 
\frac{1}{m_p \mu_e}\int_{\mathcal{C}}  n_e m_p \mu_e dV . \nonumber
\end{eqnarray}
Since the gas mass is given by an integral over a sphere $\mathcal{S}$
of radius $r_{2500}$,
\begin{eqnarray}
\Mgas = \int_{\mathcal{S}}  n_e m_p \mu_e dV \nonumber
\end{eqnarray}
the relationship between $Y$ and $Y_X$ is 
\begin{equation}
Y D_A^2 =  \left(\frac{\sigma_T}{m_e c^2}\right) \frac{1}{m_p \mu_e} C Y_X
\label{eq:Y-Yx}
\end{equation}
where the constant $C=\int_{\mathcal{C}} n_e dV / \int_{\mathcal{S}}
n_e dV$ accounts for the different domain of integration of $Y$ and
$Y_X$, and depends on the parameters of the $\beta$ model. We
calculate $C$ separately for each cluster, and typically find
$C\sim$2.

\begin{figure}
\includegraphics[width=3.8in]{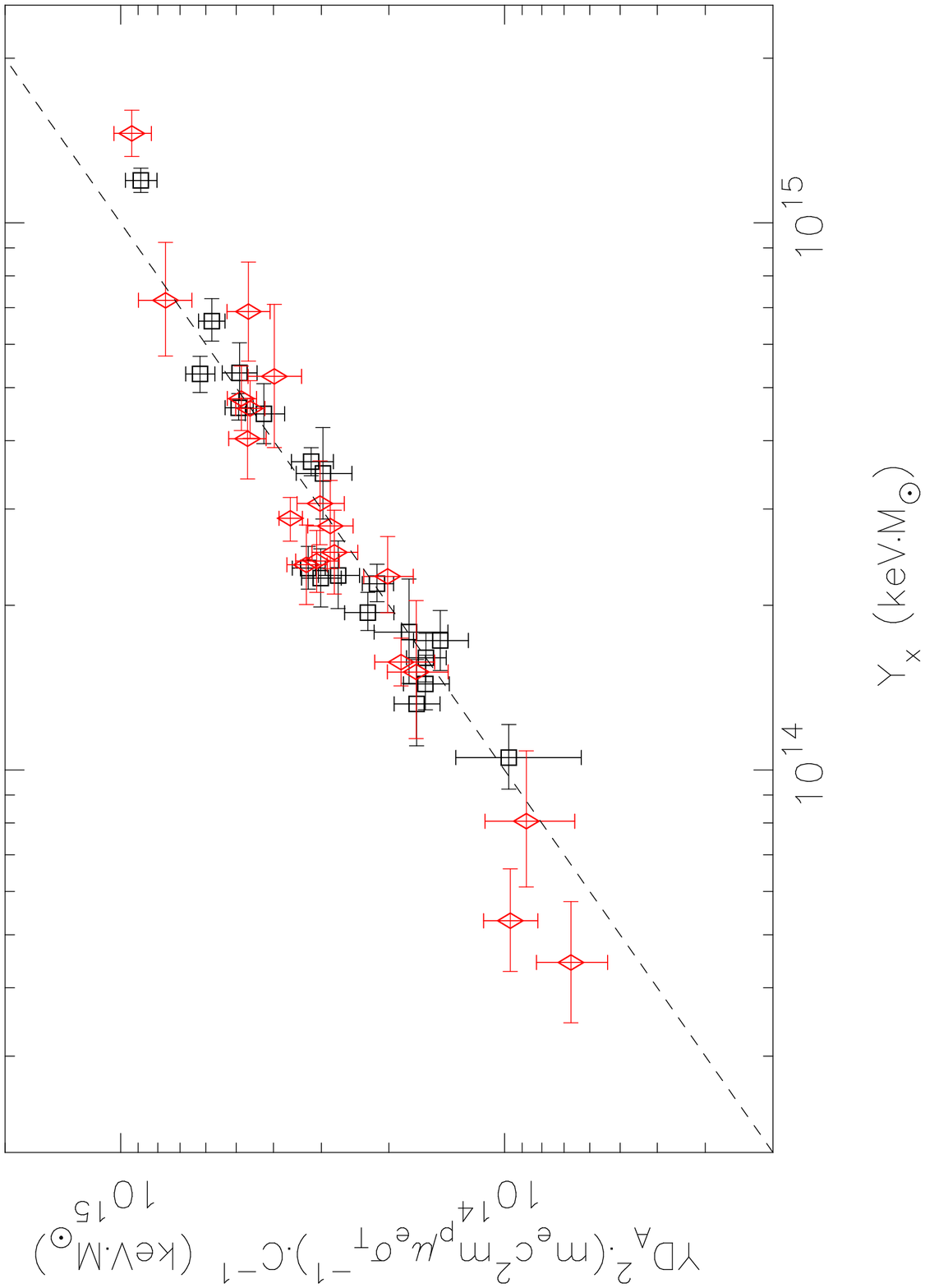}
\caption{Comparison between the \sze\ \Ysz\ parameter and the X-ray quantity $Y_X$
(see Section \ref{sz-x} for explanation of the normalization constant); 
open squares (in black) are clusters at $0.14\leq z \leq 0.30$, 
open diamonds (in
red) are clusters at $0.30 < z \leq 0.89$. The dashed black line corresponds to $y=x$.\label{Y-Yx}}
\end{figure}

In Figure \ref{Y-Yx} we plot $Y$, as derived from the joint SZE/X-ray
analysis, against $Y_X$.  
A fit of the data to the relationship in Equation \ref{eq:Y-Yx} (the 
dotted line in Figure \ref{Y-Yx}, with no degrees of freedom) 
results in an acceptable  $\chi^2$ statistic, corresponding to
a null hypothesis probability of 80.4\%.
The agreement with the relationship in
Equation \ref{eq:Y-Yx} shows that $Y_X$ is an unbiased
estimator of the integrated Compton-$y$ parameter, within the
uncertainties of the current measurements.

\section{Comparison with theoretical simulations}
\begin{figure}
\includegraphics[width=4.4in]{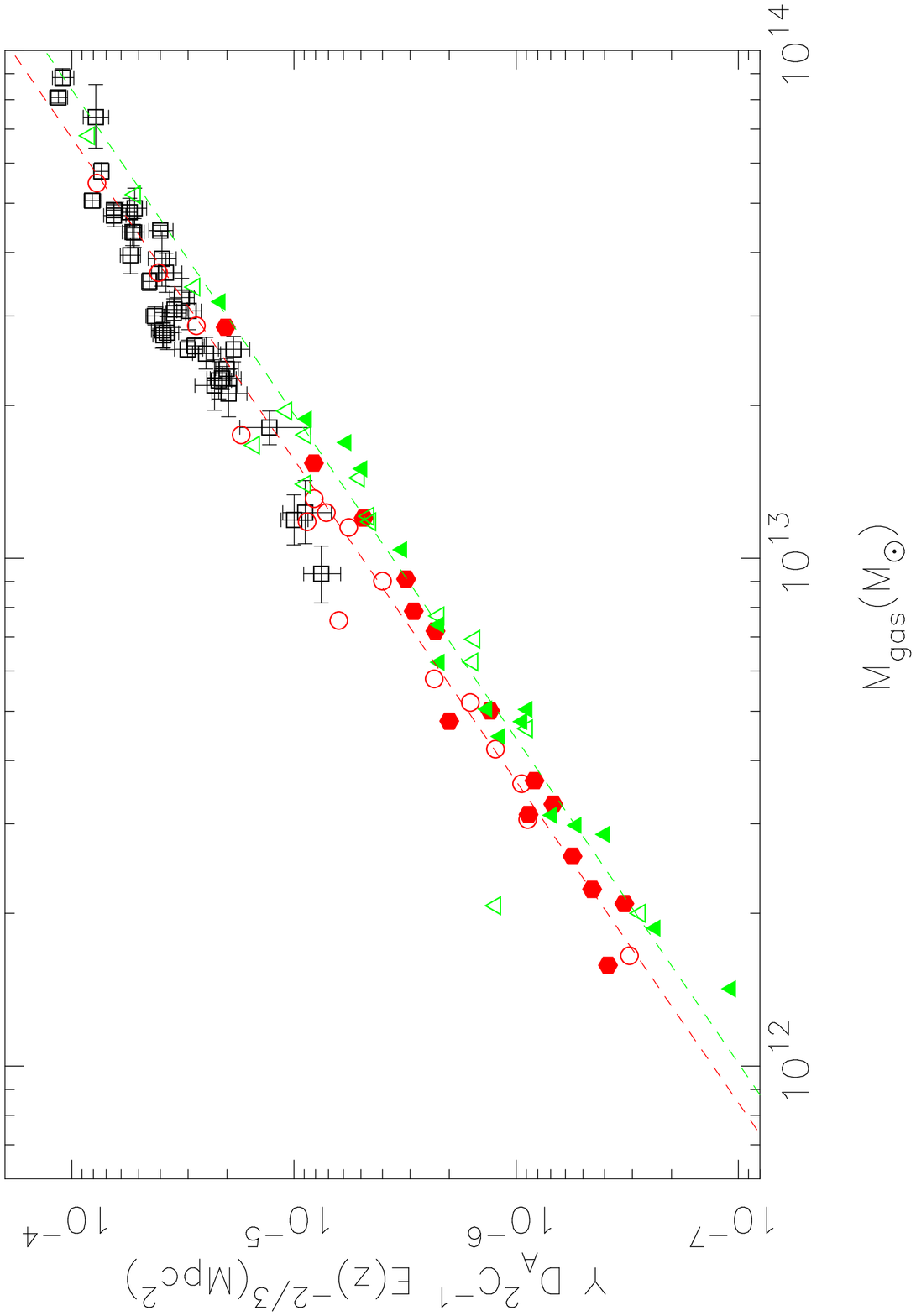}
\caption{$M_{gas}$ vs. $Y$ for simulated and observed clusters. Open
squares (in black) are OVRO/BIMA/Chandra measurements.
Also shown are simulated
clusters from a cooling and star-formation feedback model
(red circles) and simulated clusters from a non-radiative model
(green triangles) from
\citet{nagai2006}. For comparison, all of the $M_{gas}$ and $Y$ quantities are integrated over a spherical volume,
as described in Section~\ref{sz-x}.
Open symbols represent simulated clusters at $z$=0 and 
filled symbols represent simulated clusters at $z$=0.6. The
power-law fits to the simulated clusters (dashed lines) are 
$A=-25.83\pm0.52$, $B=1.58\pm 0.04$ for the cooling/star-formation model and 
$A=-25.79\pm0.80$, $B=1.56\pm 0.06$ for the non-radiative model.\label{daisuke}}
\end{figure}

We compare our results with those of recent cosmological cluster
simulations \citep{nagai2006,nagai2007b} that include radiative
cooling, UV heating, star formation, and stellar feedback processes
in addition to the standard gas dynamics.  In Figure \ref{daisuke} we
compare 16 clusters simulated at $z$=0 and 0.6 using cooling and star
formation feedback processes (in red), the same sets of clusters
performed using non-radiative gas dynamics (in green), and our 38
clusters observed with \chandra\ and OVRO-BIMA (in black). 

The best-fit power-law models that describe the two simulations
are shown as dashed lines in Figure \ref{daisuke}.
A fit of our data to 
the cooling and star formation model (red dashed line) 
results in a $\chi^2$ null hypothesis probability of 99.9\%,
and the non-radiative model (green dashed line) has a probability of 0.5\%.
The
comparison indicates that both simulation models show  a similar slope
to the observed clusters, with  the cooling and star formation
feedback model providing a better match to the data.

\section{Discussion and conclusions}

We have investigated scaling relations between the integrated
Compton-$y$ parameter \Ysz\ and total mass, gas mass, and gas
temperature using 38 clusters observed with \chandra\ and \ovro-\bima.
Fits of the \Ysz-\mgas, \Ysz-\mtot\ and \Ysz-$kT$ data to a power-law
model agree with the slope predicted by a self-similar model in which
the evolution of clusters is dominated by gravitational processes.
The normalization of the \Ysz-\mgas\ scaling relation agrees well with
the numerical simulations of \citet{nagai2007b}, in which
collisionless dynamics of dark matter and gas dynamics are
complemented by cooling, star formation and feedback phenomena.  The
agreement provides observational evidence that non-gravitational
phenomena may also be an important factor in the physics of clusters.

The redshift coverage of our sample enabled an analysis of the scaling
relations as function of redshift, by defining a low redshift sample
($0.14 \leq z \leq 0.30$, 19 clusters) and a high-redshift sample
($0.30 < z \leq 0.89$, 19 clusters).  Both samples follow the
prediction based on the self-similar model.  Our data indicate no
significant evolution in the SZE properties of clusters at redshift $z
\lsim 1$.

We also measure the cluster mass and integrated Y parameter using the
SZE data alone, without making use of the \chandra\ spectral and
spatial information.  These measurements are in good agreement with
those based on the joint X-ray/SZE analysis, providing 
evidence that SZE surveys can be used to determine the number density
of galaxy clusters as functions of mass and redshift.

\vskip 24 pt

\acknowledgments 

This work was made possible by the skill and dedication of Leon van Speybroeck and
his colleagues on the Chandra project, who constructed an exceptional
observatory and obtained deep X-ray observations for a large sample of
galaxy clusters.  We thank E.\ Leitch and the anonymous referee for 
excellent suggestions on the manuscript. 
The support of the BIMA and OVRO staff over many
years is also gratefully acknowledged, including J.R.\ Forster,
C.\ Giovanine, R.\ Lawrence, S.\ Padin, R.\ Plambeck, S.\ Scott and
D.\ Woody.  We thank C.\ Alexander, K.\ Coble, A.\ Cooray,
K.\ Dawson, L.\ Grego, G.\ Holder, W.\ Holzapfel,
A.\ Miller, J.\ Mohr, S.\ Patel , E.\ Reese, and P.\ Whitehouse for their outstanding
contributions to the SZE instrumentation, observations, and analysis.

This work was supported in part by NSF grants AST-0096913 and
AST-0604982 and by the KICP NSF Physics Frontier Center grant
PHY-0114422. 
Research at the Owens Valley Radio Observatory
and the Berkeley-Illinois-Maryland Array was supported by National
Science Foundation grants AST 99-81546 and 02-28963.  Calculations
were performed at the Space Plasma Interactive Data Analysis and
Simulation Laboratory at the Center for Space Plasma and Aeronomy
Research of the University of Alabama at Huntsville.
DN is supported by the Sherman Fairchild Postdoctoral Fellowship at
Caltech. The National Radio Astronomy Observatory is a facility of the National Science Foundation operated under cooperative agreement by Associated Universities, Inc.

\footnotesize
\bibliographystyle{apj}

\begin{deluxetable}{llcc|ccccccc}
\renewcommand{\arraystretch}{0.85}
\tabletypesize{\scriptsize}
\tablecaption{Cluster Parameters from Joint Analysis of X-ray and SZE Data\label{table:results}}
\tablehead{Cluster &z &  $D_A$ & E(z) & \multicolumn{2}{c}{$r_{2500}$} &  $kT$ & $M_{gas}$ & $M_{tot}$ & $Y$ & $f_{gas}$  \\ 
  & & (Gpc) &  & (") & (kpc) & (keV) & ($10^{13} M_{\odot}$) & ($10^{14} M_{\odot}$) & (mJy) & }
\startdata 
Abell~1413\dotfill &0.14 & $ 0.52 $ & $ 1.07 $ & $ 206\pm^{  5}_{  5}$ & $ 519\pm^{ 14}_{ 12}$ & $ 7.5\pm^{0.4}_{0.3}$ & $ 2.6\pm^{0.1}_{0.1}$ & $ 2.2\pm^{0.2}_{0.1}$ & $ 2.99\pm^{0.44}_{0.43}$ & $ 0.120\pm^{0.004}_{0.004}$ \\ 
Abell~1689\dotfill &0.18 & $ 0.63 $ & $ 1.09 $ & $ 219\pm^{  5}_{  5}$ & $ 664\pm^{ 16}_{ 17}$ & $ 10.5\pm^{0.5}_{0.5}$ & $ 5.1\pm^{0.2}_{0.2}$ & $ 5.0\pm^{0.4}_{0.4}$ & $ 3.79\pm^{0.34}_{0.31}$ & $ 0.102\pm^{0.004}_{0.004}$ \\ 
Abell~1835\dotfill &0.25 & $ 0.81 $ & $ 1.13 $ & $ 172\pm^{  5}_{  4}$ & $ 672\pm^{ 20}_{ 17}$ & $ 11.4\pm^{0.7}_{0.6}$ & $ 5.8\pm^{0.2}_{0.2}$ & $ 5.6\pm^{0.5}_{0.4}$ & $ 2.09\pm^{0.17}_{0.16}$ & $ 0.104\pm^{0.005}_{0.005}$ \\ 
Abell~1914\dotfill &0.17 & $ 0.60 $ & $ 1.09 $ & $ 228\pm^{  5}_{  4}$ & $ 660\pm^{ 13}_{ 12}$ & $ 9.5\pm^{0.4}_{0.3}$ & $ 4.8\pm^{0.1}_{0.1}$ & $ 4.8\pm^{0.3}_{0.2}$ & $ 3.01\pm^{0.25}_{0.25}$ & $ 0.101\pm^{0.003}_{0.004}$ \\ 
Abell~1995\dotfill &0.32 & $ 0.96 $ & $ 1.18 $ & $ 133\pm^{  5}_{  5}$ & $ 621\pm^{ 21}_{ 22}$ & $ 8.2\pm^{0.4}_{0.4}$ & $ 3.5\pm^{0.1}_{0.1}$ & $ 4.7\pm^{0.5}_{0.5}$ & $ 0.75\pm^{0.05}_{0.05}$ & $ 0.074\pm^{0.005}_{0.005}$ \\ 
Abell~2111\dotfill &0.23 & $ 0.76 $ & $ 1.12 $ & $ 141\pm^{ 10}_{  9}$ & $ 518\pm^{ 36}_{ 33}$ & $ 8.2\pm^{1.0}_{0.8}$ & $ 2.2\pm^{0.3}_{0.2}$ & $ 2.5\pm^{0.6}_{0.4}$ & $ 0.95\pm^{0.21}_{0.20}$ & $ 0.088\pm^{0.008}_{0.008}$ \\ 
Abell~2163\dotfill &0.20 & $ 0.68 $ & $ 1.10 $ & $ 206\pm^{  3}_{  3}$ & $ 682\pm^{ 10}_{ 10}$ & $ 14.8\pm^{0.4}_{0.4}$ & $ 8.1\pm^{0.2}_{0.2}$ & $ 5.5\pm^{0.2}_{0.2}$ & $ 6.89\pm^{0.66}_{0.63}$ & $ 0.147\pm^{0.003}_{0.003}$ \\ 
Abell~2204\dotfill &0.15 & $ 0.54 $ & $ 1.08 $ & $ 256\pm^{ 13}_{ 11}$ & $ 671\pm^{ 34}_{ 30}$ & $ 11.2\pm^{0.8}_{0.7}$ & $ 4.7\pm^{0.3}_{0.2}$ & $ 5.0\pm^{0.8}_{0.6}$ & $ 4.43\pm^{0.50}_{0.52}$ & $ 0.096\pm^{0.009}_{0.010}$ \\ 
Abell~2218\dotfill &0.18 & $ 0.63 $ & $ 1.09 $ & $ 191\pm^{  6}_{  5}$ & $ 581\pm^{ 18}_{ 16}$ & $ 7.8\pm^{0.4}_{0.4}$ & $ 3.0\pm^{0.1}_{0.1}$ & $ 3.3\pm^{0.3}_{0.3}$ & $ 1.94\pm^{0.19}_{0.18}$ & $ 0.090\pm^{0.004}_{0.004}$ \\ 
Abell~2259\dotfill &0.16 & $ 0.57 $ & $ 1.08 $ & $ 172\pm^{  8}_{  8}$ & $ 476\pm^{ 23}_{ 22}$ & $ 5.8\pm^{0.4}_{0.4}$ & $ 1.8\pm^{0.1}_{0.1}$ & $ 1.8\pm^{0.3}_{0.2}$ & $ 0.82\pm^{0.30}_{0.29}$ & $ 0.101\pm^{0.007}_{0.007}$ \\ 
Abell~2261\dotfill &0.22 & $ 0.73 $ & $ 1.12 $ & $ 148\pm^{  7}_{  6}$ & $ 525\pm^{ 24}_{ 22}$ & $ 7.4\pm^{0.6}_{0.5}$ & $ 3.0\pm^{0.2}_{0.2}$ & $ 2.6\pm^{0.4}_{0.3}$ & $ 1.34\pm^{0.16}_{0.16}$ & $ 0.119\pm^{0.008}_{0.008}$ \\ 
Abell~267\dotfill &0.23 & $ 0.76 $ & $ 1.12 $ & $ 132\pm^{  9}_{  7}$ & $ 484\pm^{ 31}_{ 28}$ & $ 5.9\pm^{0.7}_{0.5}$ & $ 2.2\pm^{0.2}_{0.2}$ & $ 2.0\pm^{0.4}_{0.3}$ & $ 0.72\pm^{0.10}_{0.09}$ & $ 0.110\pm^{0.011}_{0.011}$ \\ 
Abell~370\dotfill &0.38 & $ 1.07 $ & $ 1.22 $ & $  97\pm^{  4}_{  4}$ & $ 508\pm^{ 21}_{ 21}$ & $ 8.7\pm^{0.5}_{0.5}$ & $ 2.8\pm^{0.2}_{0.2}$ & $ 2.8\pm^{0.4}_{0.3}$ & $ 0.71\pm^{0.09}_{0.08}$ & $ 0.100\pm^{0.005}_{0.005}$ \\ 
Abell~586\dotfill &0.17 & $ 0.60 $ & $ 1.09 $ & $ 182\pm^{  8}_{  7}$ & $ 529\pm^{ 23}_{ 20}$ & $ 6.4\pm^{0.5}_{0.4}$ & $ 2.3\pm^{0.1}_{0.1}$ & $ 2.5\pm^{0.3}_{0.3}$ & $ 1.03\pm^{0.14}_{0.14}$ & $ 0.091\pm^{0.007}_{0.007}$ \\ 
Abell~611\dotfill &0.29 & $ 0.90 $ & $ 1.16 $ & $ 111\pm^{  4}_{  3}$ & $ 482\pm^{ 16}_{ 15}$ & $ 6.8\pm^{0.4}_{0.4}$ & $ 2.4\pm^{0.1}_{0.1}$ & $ 2.1\pm^{0.2}_{0.2}$ & $ 0.54\pm^{0.06}_{0.06}$ & $ 0.111\pm^{0.006}_{0.006}$ \\ 
Abell~665\dotfill &0.18 & $ 0.63 $ & $ 1.09 $ & $ 162\pm^{  4}_{  3}$ & $ 490\pm^{ 11}_{ 10}$ & $ 8.4\pm^{0.4}_{0.3}$ & $ 2.6\pm^{0.1}_{0.1}$ & $ 2.0\pm^{0.1}_{0.1}$ & $ 2.68\pm^{0.24}_{0.25}$ & $ 0.131\pm^{0.004}_{0.004}$ \\ 
Abell~68\dotfill &0.26 & $ 0.83 $ & $ 1.14 $ & $ 153\pm^{ 10}_{  9}$ & $ 616\pm^{ 40}_{ 37}$ & $ 9.6\pm^{1.1}_{1.0}$ & $ 3.6\pm^{0.3}_{0.3}$ & $ 4.3\pm^{0.9}_{0.7}$ & $ 1.01\pm^{0.16}_{0.15}$ & $ 0.084\pm^{0.009}_{0.008}$ \\ 
Abell~697\dotfill &0.28 & $ 0.88 $ & $ 1.15 $ & $ 134\pm^{  5}_{  5}$ & $ 568\pm^{ 21}_{ 21}$ & $ 10.2\pm^{0.7}_{0.6}$ & $ 4.4\pm^{0.3}_{0.3}$ & $ 3.5\pm^{0.4}_{0.4}$ & $ 1.67\pm^{0.19}_{0.19}$ & $ 0.126\pm^{0.007}_{0.006}$ \\ 
Abell~773\dotfill &0.22 & $ 0.73 $ & $ 1.12 $ & $ 148\pm^{  6}_{  5}$ & $ 527\pm^{ 20}_{ 19}$ & $ 8.2\pm^{0.6}_{0.5}$ & $ 2.7\pm^{0.2}_{0.2}$ & $ 2.6\pm^{0.3}_{0.3}$ & $ 1.68\pm^{0.19}_{0.19}$ & $ 0.106\pm^{0.006}_{0.005}$ \\ 
CL~J0016+1609\dotfill &0.54 & $ 1.31 $ & $ 1.34 $ & $  80\pm^{  3}_{  3}$ & $ 507\pm^{ 19}_{ 19}$ & $ 10.5\pm^{0.6}_{0.6}$ & $ 4.4\pm^{0.3}_{0.3}$ & $ 3.3\pm^{0.4}_{0.4}$ & $ 0.73\pm^{0.06}_{0.06}$ & $ 0.131\pm^{0.007}_{0.006}$ \\ 
CL~J1226+3332\dotfill &0.89 & $ 1.60 $ & $ 1.65 $ & $  66\pm^{  7}_{  6}$ & $ 512\pm^{ 58}_{ 50}$ & $ 13.5\pm^{2.7}_{2.2}$ & $ 3.9\pm^{0.5}_{0.5}$ & $ 5.2\pm^{2.0}_{1.4}$ & $ 0.35\pm^{0.05}_{0.05}$ & $ 0.075\pm^{0.015}_{0.014}$ \\ 
MACS~J0647.7+7015\dotfill &0.58 & $ 1.36 $ & $ 1.37 $ & $  92\pm^{  6}_{  6}$ & $ 606\pm^{ 41}_{ 38}$ & $ 14.1\pm^{1.8}_{1.6}$ & $ 4.9\pm^{0.5}_{0.4}$ & $ 6.0\pm^{1.3}_{1.1}$ & $ 0.62\pm^{0.08}_{0.07}$ & $ 0.082\pm^{0.009}_{0.008}$ \\ 
MACS~J0744.8+3927\dotfill &0.69 & $ 1.47 $ & $ 1.47 $ & $  59\pm^{  3}_{  3}$ & $ 420\pm^{ 25}_{ 23}$ & $ 8.1\pm^{0.8}_{0.7}$ & $ 3.1\pm^{0.3}_{0.2}$ & $ 2.3\pm^{0.4}_{0.4}$ & $ 0.34\pm^{0.04}_{0.04}$ & $ 0.136\pm^{0.012}_{0.011}$ \\ 
MACS~J1149.5+2223\dotfill &0.54 & $ 1.31 $ & $ 1.34 $ & $  71\pm^{  4}_{  4}$ & $ 449\pm^{ 25}_{ 23}$ & $ 9.9\pm^{0.8}_{0.7}$ & $ 3.1\pm^{0.3}_{0.3}$ & $ 2.3\pm^{0.4}_{0.3}$ & $ 0.58\pm^{0.08}_{0.07}$ & $ 0.134\pm^{0.008}_{0.008}$ \\ 
MACS~J1311.0-0310\dotfill &0.49 & $ 1.25 $ & $ 1.30 $ & $  74\pm^{  8}_{  7}$ & $ 448\pm^{ 46}_{ 40}$ & $ 7.2\pm^{1.5}_{1.1}$ & $ 2.1\pm^{0.2}_{0.2}$ & $ 2.2\pm^{0.7}_{0.5}$ & $ 0.28\pm^{0.05}_{0.05}$ & $ 0.097\pm^{0.019}_{0.017}$ \\ 
MACS~J1423.8+2404\dotfill &0.55 & $ 1.32 $ & $ 1.35 $ & $  66\pm^{  2}_{  2}$ & $ 422\pm^{ 14}_{ 13}$ & $ 7.0\pm^{0.4}_{0.4}$ & $ 2.3\pm^{0.1}_{0.1}$ & $ 1.9\pm^{0.2}_{0.2}$ & $ 0.28\pm^{0.05}_{0.05}$ & $ 0.116\pm^{0.006}_{0.006}$ \\ 
MACS~J2129.4-0741\dotfill &0.57 & $ 1.35 $ & $ 1.36 $ & $  73\pm^{  5}_{  4}$ & $ 474\pm^{ 30}_{ 27}$ & $ 8.6\pm^{1.0}_{0.8}$ & $ 3.3\pm^{0.3}_{0.3}$ & $ 2.8\pm^{0.6}_{0.5}$ & $ 0.39\pm^{0.05}_{0.05}$ & $ 0.116\pm^{0.011}_{0.011}$ \\ 
MACS~J2214.9-1359\dotfill &0.48 & $ 1.23 $ & $ 1.29 $ & $  91\pm^{  5}_{  5}$ & $ 547\pm^{ 30}_{ 29}$ & $ 10.2\pm^{1.0}_{0.9}$ & $ 3.9\pm^{0.3}_{0.3}$ & $ 3.9\pm^{0.7}_{0.6}$ & $ 0.77\pm^{0.09}_{0.08}$ & $ 0.102\pm^{0.009}_{0.008}$ \\ 
MACS~J2228.5+2036\dotfill &0.41 & $ 1.12 $ & $ 1.24 $ & $  81\pm^{  4}_{  4}$ & $ 444\pm^{ 22}_{ 20}$ & $ 8.4\pm^{0.8}_{0.7}$ & $ 2.8\pm^{0.2}_{0.2}$ & $ 2.0\pm^{0.3}_{0.3}$ & $ 0.94\pm^{0.11}_{0.11}$ & $ 0.138\pm^{0.009}_{0.009}$ \\ 
MS~0451.6-0305\dotfill &0.55 & $ 1.32 $ & $ 1.35 $ & $  82\pm^{  4}_{  3}$ & $ 526\pm^{ 23}_{ 22}$ & $ 9.9\pm^{0.8}_{0.7}$ & $ 4.8\pm^{0.3}_{0.3}$ & $ 3.8\pm^{0.5}_{0.5}$ & $ 0.66\pm^{0.05}_{0.05}$ & $ 0.128\pm^{0.009}_{0.009}$ \\ 
MS~1054.5-0321\dotfill &0.83 & $ 1.57 $ & $ 1.59 $ & $  89\pm^{  8}_{  7}$ & $686\pm^{106}_{198}$ & $ 9.8\pm^{1.1}_{0.9}$ & $ 7.4\pm^{1.2}_{1.0}$ & $ 4.5\pm^{1.4}_{1.0}$ & $ 0.77\pm^{0.11}_{0.10}$ & $ 0.164\pm^{0.019}_{0.019}$ \\
MS~1137.5+6625\dotfill &0.78 & $ 1.54 $ & $ 1.55 $ & $  42\pm^{  3}_{  3}$ & $ 311\pm^{ 25}_{ 22}$ & $ 4.5\pm^{0.5}_{0.4}$ & $ 1.2\pm^{0.1}_{0.1}$ & $ 1.0\pm^{0.3}_{0.2}$ & $ 0.09\pm^{0.01}_{0.01}$ & $ 0.115\pm^{0.014}_{0.013}$ \\ 
MS~1358.4+6245\dotfill &0.33 & $ 0.98 $ & $ 1.19 $ & $ 113\pm^{  6}_{  5}$ & $ 539\pm^{ 28}_{ 25}$ & $ 8.9\pm^{0.9}_{0.7}$ & $ 2.5\pm^{0.2}_{0.2}$ & $ 3.1\pm^{0.5}_{0.4}$ & $ 0.56\pm^{0.08}_{0.08}$ & $ 0.081\pm^{0.006}_{0.006}$ \\ 
MS~2053.7-0449\dotfill &0.58 & $ 1.36 $ & $ 1.37 $ & $  54\pm^{  5}_{  5}$ & $ 358\pm^{ 34}_{ 30}$ & $ 4.8\pm^{0.7}_{0.6}$ & $ 0.9\pm^{0.1}_{0.1}$ & $ 1.2\pm^{0.4}_{0.3}$ & $ 0.09\pm^{0.02}_{0.02}$ & $ 0.076\pm^{0.012}_{0.011}$ \\ 
RX~J1347.5-1145\dotfill &0.45 & $ 1.19 $ & $ 1.27 $ & $ 122\pm^{  4}_{  4}$ & $ 706\pm^{ 22}_{ 21}$ & $ 16.5\pm^{1.0}_{0.9}$ & $ 8.8\pm^{0.4}_{0.3}$ & $ 8.1\pm^{0.8}_{0.7}$ & $ 1.62\pm^{0.18}_{0.18}$ & $ 0.109\pm^{0.006}_{0.005}$ \\ 
RX~J1716.4+6708\dotfill &0.81 & $ 1.56 $ & $ 1.57 $ & $  45\pm^{  4}_{  4}$ & $ 341\pm^{ 33}_{ 29}$ & $ 6.6\pm^{1.1}_{0.9}$ & $ 1.2\pm^{0.2}_{0.2}$ & $ 1.4\pm^{0.4}_{0.3}$ & $ 0.10\pm^{0.03}_{0.02}$ & $ 0.088\pm^{0.013}_{0.011}$ \\ 
RX~J2129.7+0005\dotfill &0.24 & $ 0.78 $ & $ 1.13 $ & $ 128\pm^{  5}_{  5}$ & $ 486\pm^{ 20}_{ 19}$ & $ 6.7\pm^{0.5}_{0.5}$ & $ 2.6\pm^{0.2}_{0.1}$ & $ 2.1\pm^{0.3}_{0.2}$ & $ 0.66\pm^{0.11}_{0.10}$ & $ 0.124\pm^{0.008}_{0.008}$ \\ 
ZW~3146\dotfill &0.29 & $ 0.90 $ & $ 1.16 $ & $ 132\pm^{  3}_{  3}$ & $ 574\pm^{ 11}_{ 11}$ & $ 8.3\pm^{0.3}_{0.3}$ & $ 4.4\pm^{0.1}_{0.1}$ & $ 3.6\pm^{0.2}_{0.2}$ & $ 0.88\pm^{0.11}_{0.11}$ & $ 0.122\pm^{0.004}_{0.004}$ \\ 
\enddata 
\end{deluxetable}

\begin{deluxetable}{lcccccc}
\tabletypesize{\small}
\tablecaption{Scaling Relations from 
Joint Analysis of X-ray and SZE Data\label{table:scaling}}
\tablehead{Scaling&\multicolumn{2}{c}{All
    clusters}&\multicolumn{2}{c}{0.14$\leq z
    \leq$0.30}&\multicolumn{2}{c}{0.31$< z \leq$0.89}\\
relation & \multicolumn{2}{c}{\hrulefill} & \multicolumn{2}{c}{\hrulefill} & \multicolumn{2}{c}{\hrulefill}\\
    & $A$ & $B$ &  $A$ & $B$ & $A$ & $B$}
\startdata
\fgas,\mgas & -2.86$\pm$1.09 & 0.14$\pm$0.08 & -2.60$\pm$1.79  & 0.12$\pm$0.13 &-3.00$\pm$1.37  & 0.15$\pm$0.10 \\
\Ysz,$kT$ & -6.24$\pm$0.22  &  2.37$\pm$0.23 &-6.33$\pm$0.32  &  2.46$\pm$0.34 & -6.13$\pm$0.30  &  2.27$\pm$0.30\\
\Ysz,\mgas & -23.25$\pm$1.77 & 1.41$\pm$0.13 & -25.86$\pm$3.45 & 1.60$\pm$0.25 & -21.43$\pm$3.00 & 1.28$\pm$0.15\\
\Ysz,\Mtot & -28.23$\pm$3.00  &  1.66$\pm$0.20&-31.20$\pm$5.35  &  1.87$\pm$0.35 & -25.45$\pm$3.46  &  1.47$\pm$0.23\\

\enddata
\end{deluxetable}

\begin{deluxetable}{lc|ccccc}
\renewcommand{\arraystretch}{0.85}
\tabletypesize{\footnotesize}
\tablecaption{Cluster Parameters from Analysis of SZE Data\label{table:sz-results}}
\tablehead{Cluster & $z$ & $r_{2500}$ &  $kT$ & $M_{gas}$ & $M_{tot}$ & $Y$   \\ 
    & & (") & (keV) & ($10^{13}\; M_{\odot}$) & ($10^{14}\; M_{\odot}$) & ($10^{-10}$) }
\startdata 
Abell~1689\dotfill &0.18 & $ 196\pm^{  8}_{  8}$ & $ 8.0\pm^{0.8}_{0.7}$ & $ 4.2\pm^{0.6}_{0.5}$ & $ 3.6\pm^{0.5}_{0.4}$ & $ 1.88\pm^{0.49}_{0.38}$ \\ 
Abell~1835\dotfill &0.25 & $ 169\pm^{  5}_{  5}$ & $ 11.4\pm^{1.0}_{0.9}$ & $ 6.1\pm^{0.6}_{0.6}$ & $ 5.3\pm^{0.5}_{0.5}$ & $ 2.66\pm^{0.60}_{0.48}$ \\ 
Abell~1914\dotfill &0.17 & $ 212\pm^{ 10}_{  9}$ & $ 8.5\pm^{1.0}_{0.8}$ & $ 4.5\pm^{0.7}_{0.6}$ & $ 3.9\pm^{0.6}_{0.5}$ & $ 2.39\pm^{0.77}_{0.54}$ \\ 
Abell~1995\dotfill &0.32 & $ 117\pm^{  3}_{  3}$ & $ 8.4\pm^{0.6}_{0.6}$ & $ 3.7\pm^{0.3}_{0.3}$ & $ 3.2\pm^{0.3}_{0.3}$ & $ 0.84\pm^{0.16}_{0.14}$ \\ 
Abell~2111\dotfill &0.23 & $ 124\pm^{  9}_{  9}$ & $ 5.4\pm^{1.6}_{1.0}$ & $ 2.0\pm^{0.5}_{0.4}$ & $ 1.7\pm^{0.4}_{0.3}$ & $ 0.45\pm^{0.36}_{0.18}$ \\ 
Abell~2163\dotfill &0.20 & $ 229\pm^{ 13}_{ 12}$ & $ 15.6\pm^{2.4}_{2.0}$ & $ 8.7\pm^{1.6}_{1.3}$ & $ 7.5\pm^{1.4}_{1.2}$ & $ 8.03\pm^{3.16}_{2.26}$ \\ 
Abell~2218\dotfill &0.18 & $ 206\pm^{ 12}_{ 11}$ & $ 10.0\pm^{1.5}_{1.3}$ & $ 4.8\pm^{0.9}_{0.8}$ & $ 4.1\pm^{0.8}_{0.6}$ & $ 3.20\pm^{1.24}_{0.90}$ \\ 
Abell~2261\dotfill &0.22 & $ 146\pm^{ 10}_{  8}$ & $ 6.4\pm^{1.0}_{0.8}$ & $ 2.9\pm^{0.6}_{0.5}$ & $ 2.5\pm^{0.5}_{0.4}$ & $ 0.76\pm^{0.34}_{0.21}$ \\ 
Abell~267\dotfill &0.23 & $ 138\pm^{  7}_{  7}$ & $ 6.5\pm^{1.0}_{0.8}$ & $ 2.7\pm^{0.4}_{0.4}$ & $ 2.3\pm^{0.4}_{0.3}$ & $ 0.75\pm^{0.30}_{0.21}$ \\ 
Abell~370\dotfill &0.38 & $  96\pm^{  4}_{  3}$ & $ 7.3\pm^{0.8}_{0.7}$ & $ 3.0\pm^{0.3}_{0.3}$ & $ 2.6\pm^{0.3}_{0.3}$ & $ 0.45\pm^{0.13}_{0.10}$ \\ 
Abell~665\dotfill &0.18 & $ 181\pm^{ 12}_{ 11}$ & $ 7.4\pm^{1.6}_{1.1}$ & $ 3.3\pm^{0.7}_{0.6}$ & $ 2.8\pm^{0.6}_{0.5}$ & $ 1.55\pm^{0.89}_{0.51}$ \\ 
Abell~697\dotfill &0.28 & $ 130\pm^{ 13}_{ 10}$ & $ 8.3\pm^{2.2}_{1.5}$ & $ 3.7\pm^{1.2}_{0.8}$ & $ 3.2\pm^{1.0}_{0.7}$ & $ 0.99\pm^{0.75}_{0.38}$ \\ 
Abell~773\dotfill &0.22 & $ 150\pm^{  9}_{  8}$ & $ 6.9\pm^{1.0}_{0.8}$ & $ 3.1\pm^{0.6}_{0.5}$ & $ 2.7\pm^{0.5}_{0.4}$ & $ 0.93\pm^{0.38}_{0.26}$ \\ 
CL~J0016+1609\dotfill &0.54 & $  81\pm^{  2}_{  3}$ & $ 12.6\pm^{0.9}_{0.9}$ & $ 4.1\pm^{0.4}_{0.5}$ & $ 3.5\pm^{0.3}_{0.4}$ & $ 0.92\pm^{0.11}_{0.11}$ \\ 
CL~J1226+3332\dotfill &0.89 & $  53\pm^{  1}_{  2}$ & $ 9.2\pm^{0.5}_{0.5}$ & $ 3.2\pm^{0.3}_{0.3}$ & $ 2.7\pm^{0.2}_{0.3}$ & $ 0.27\pm^{0.04}_{0.03}$ \\ 
MACS~J0647.7+7015\dotfill &0.58 & $  72\pm^{  3}_{  3}$ & $ 8.6\pm^{1.1}_{0.9}$ & $ 3.4\pm^{0.4}_{0.4}$ & $ 2.9\pm^{0.3}_{0.3}$ & $ 0.38\pm^{0.12}_{0.09}$ \\ 
MACS~J1311.0-0310\dotfill &0.49 & $  73\pm^{  4}_{  5}$ & $ 6.8\pm^{0.8}_{0.8}$ & $ 2.4\pm^{0.4}_{0.5}$ & $ 2.1\pm^{0.4}_{0.4}$ & $ 0.26\pm^{0.08}_{0.07}$ \\ 
MACS~J2214.9-1359\dotfill &0.48 & $  91\pm^{  2}_{  2}$ & $ 10.2\pm^{0.7}_{0.7}$ & $ 4.5\pm^{0.3}_{0.3}$ & $ 3.9\pm^{0.3}_{0.3}$ & $ 0.74\pm^{0.13}_{0.12}$ \\ 
MACS~J2228.5+2036\dotfill &0.41 & $ 100\pm^{  3}_{  4}$ & $ 11.0\pm^{1.9}_{1.4}$ & $ 4.3\pm^{0.5}_{0.5}$ & $ 3.7\pm^{0.4}_{0.4}$ & $ 1.03\pm^{0.37}_{0.27}$ \\ 
MS~0451.6-0305\dotfill &0.55 & $  84\pm^{  3}_{  3}$ & $ 11.3\pm^{1.2}_{1.0}$ & $ 4.7\pm^{0.4}_{0.4}$ & $ 4.1\pm^{0.4}_{0.4}$ & $ 0.79\pm^{0.21}_{0.16}$ \\ 
MS~1137.5+6625\dotfill &0.78 & $  49\pm^{  2}_{  3}$ & $ 6.4\pm^{0.4}_{0.4}$ & $ 1.9\pm^{0.2}_{0.3}$ & $ 1.6\pm^{0.2}_{0.3}$ & $ 0.12\pm^{0.02}_{0.02}$ \\ 
MS~1358.4+6245\dotfill &0.33 & $  99\pm^{  4}_{  5}$ & $ 6.4\pm^{0.9}_{0.6}$ & $ 2.4\pm^{0.3}_{0.3}$ & $ 2.1\pm^{0.3}_{0.3}$ & $ 0.40\pm^{0.12}_{0.09}$ \\ 
RX~J1347.5-1145\dotfill &0.45 & $ 108\pm^{  4}_{  4}$ & $ 12.7\pm^{1.1}_{1.0}$ & $ 6.5\pm^{0.7}_{0.6}$ & $ 5.6\pm^{0.6}_{0.6}$ & $ 1.41\pm^{0.32}_{0.27}$ \\ 
RX~J2129.7+0005\dotfill &0.24 & $ 123\pm^{  9}_{ 24}$ & $ 7.9\pm^{3.9}_{2.1}$ & $ 2.1\pm^{0.5}_{1.0}$ & $ 1.8\pm^{0.4}_{0.9}$ & $ 0.80\pm^{0.56}_{0.35}$ \\ 
ZW~3146\dotfill &0.29 & $ 128\pm^{  6}_{  6}$ & $ 8.3\pm^{1.0}_{0.9}$ & $ 3.9\pm^{0.6}_{0.5}$ & $ 3.3\pm^{0.5}_{0.4}$ & $ 0.94\pm^{0.31}_{0.24}$ \\ 
\enddata 
\end{deluxetable}

\begin{figure}[!h]
\includegraphics[width=4.0in]{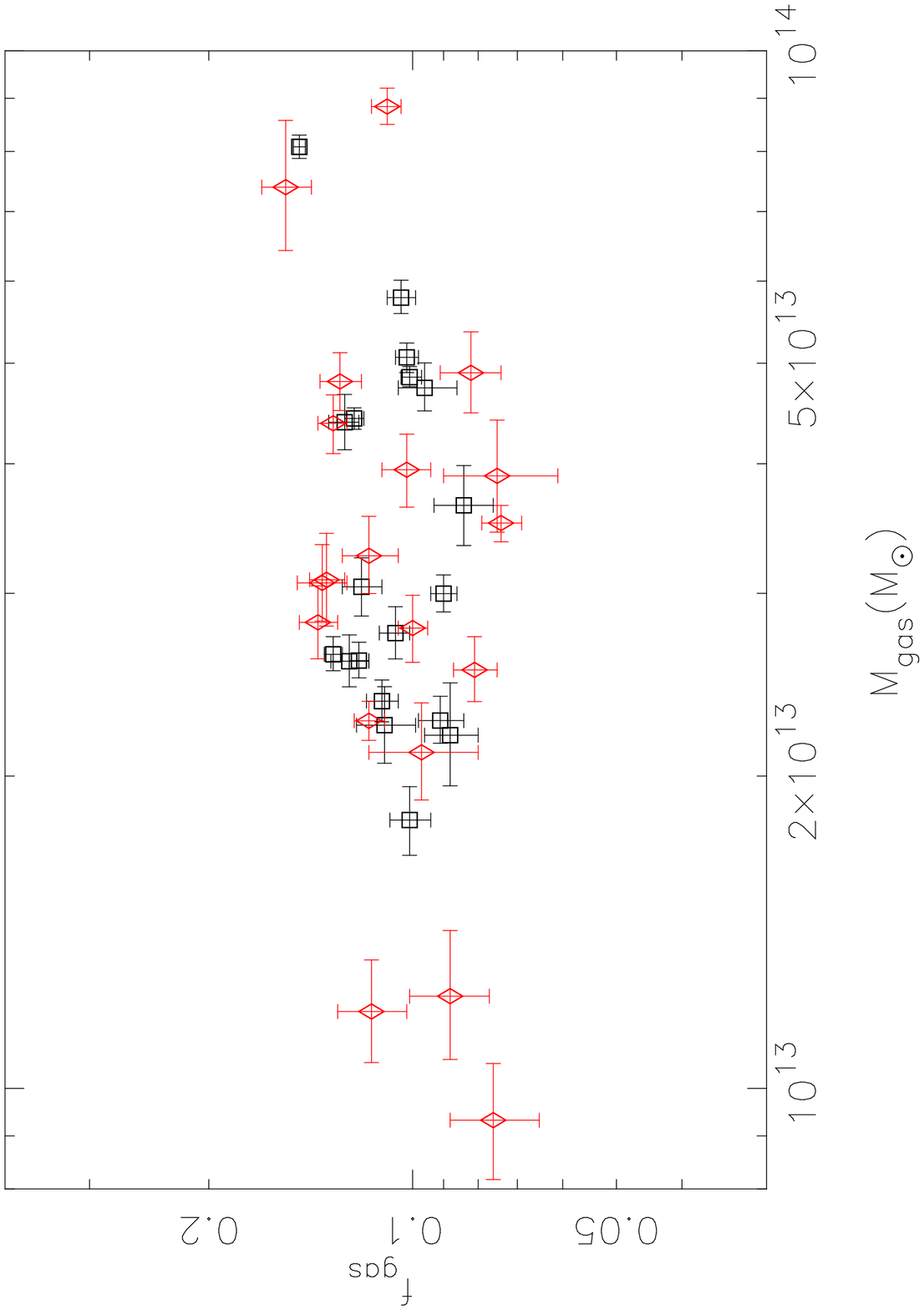}
\caption{Dependence of \fgas\ on \mgas; open squares (in black) are clusters at $0.14\leq z \leq 0.30$, 
open diamonds (in red) 
are clusters at $0.30 < z \leq 0.89$. The gas fraction at \rtfh\ shows no evidence
of evolution with mass for the
clusters in this sample.
\label{fig:fgas}}
\end{figure}

\end{document}